\documentclass[12pt]{article}
\title{ Semi classical modeling of isotropic Non-Heisenberg magnets for spin $S=1$ and linear quadrupole excitation dynamics}
\author{ Yousef Yousefi,  Khikmat Kh. Muminov  \\
Physical-Technical Institute named after S.U.Umarov\\
 Academy of Sciences of Republic of Tajikistan\\
Aini Ave 299/1, Dushanbe, Tajikistan\\
E-mail:  yousof54@yahoo.com }
\date{}
\begin{document}
\maketitle
\begin{abstract}
In this paper, equations describing one-dimensional Non-Heisenberg model are studied by use of generalized coherent states in  real parameterization and then dissipative spin wave equation for dipole and quadrupole branches is obtained if there is a small linear excitation from the ground state. Finally, it is shown that for such exchange-isotropy Hamiltonians,  optical branch of spin wave is non-dissipative.
\end{abstract}

\section{Introduction}

Many condensed matter systems fully described by use of effective continuum field models. Topologically nontrivial field configurations have an important role in modeling of systems with reduced spatial dimensionality [1]. Magnetic systems are usually modeled with the help of the Heisenberg exchange interaction [2,3]. 

However, for spin $ S>1/2$,  the general isotropic exchange goes beyond the purely Heisenberg interaction bilinear in spin operators $ \vec S_i$ and includes higher-order terms of the type  $ (\vec S_i \vec S_j)^n$  with n up to 2S [4]. Due to the spin states, the  2S+1 complex parameters are necessary  to describe each of them and this corresponds with the 4S+2 degrees of freedom. Two degree of freedom are omitted, one because of normalization condition and the other for arbitrary phase decrease, hence 4S parameters are required to completely modeled the remainder 4S degrees of freedom of spin states. [5]

 Particularly, In case S=1 with the isotropic nearest – neighbor exchange on a lattice, is derived by use of the Hamiltonian

\begin{eqnarray}
\hat H=-\sum_i J(\vec{ \hat S_i} \vec{\hat S_{i+1}})+K(\vec{ \hat S_i} \vec{\hat S_{i+1}})^2
\end{eqnarray}

Here $ \hat S_i^x , \hat S_i^y ,\hat S_i^z$   are the spin operators acting at a site i, J and K are respectively the bilinear (Heisenberg) and biquadratic exchange integral. The model (1) has been discussed recently in connection with S = 1 bosonic gases in optical lattices [6] and in the context of the deconfined quantum criticality [7,8]. Hamiltonian (1) is a special form presented in reference [9] and because of importance of quadrupole excitation in ferromagnetic Martials, it is considered here. This article does not consider the anti-ferromagnetic and nematic states.

Considering the effects of both dipole and quadrupole branches gives a nonlinear approximation. If higher order multipole effects considered, the approximation is more accurate but at the same time, deriving the equations are too complicated. In this paper, only effect of quadrupole branch for Hamiltonians described by equation (1) is considered. Study of isotropic and anisotropic spin Hamiltonian with non-Heisenberg terms are complicated due to quadrupole excitation dynamics[4,10,11]. 
Antiferromagnetic property of this excitation in states near the ground, proves the existence of it and Dzyaloshinskii calculated the effect of this excitation [12]. Also, numerical calculations more accurately justify laboratory results if the effect of quadrupole excitation in nanoparticles $Fe_8$  and $Mn_{12}$ is considered[13,14]. In addition, this method may be promising for description of the multi-spin configuration of the $Fe$ in the different ligand coordinations [15].

To calculate effect of quadrupole excitation,  at first it is necessary to obtain classical equivalent of Hamiltonian (1) and then to find out the solution of spin wave, it is necessary to analyze resulted equations in case there is small linear excitation from the ground states. Therefore, the stages of process are:

1.	Obtaining coherent states for spin s=1 which are coherent states of SU(3) group. 

2.	Calculation of average values of spin operator.

3.	Classical spin Hamiltonian equation is obtained using previously calculated values.

4.	Calculating Lagrangian equation by use of Feynman path integral over coherent states and then computing classical equations of motion.

5.	 It is necessary to substitute resulted Hamiltonian in classical equations of motion to obtain nonlinear equations of magnets. Solutions of these nonlinear equations result in soliton description of magnet that is not interested here.

6.	 Now, ground states of magnets calculated and then nonlinear equations are linearized above the ground states for small linear excitation. 

7.	Finally, spin wave equations and dispersion equations must be calculated.

In this article, we write coherent states in real parameters, because each parameter in this representation related to one degree of freedom. In complex parameterization, each parameter related to two or more degrees of freedom. Then in physical problems, the first representation is very helpful. 

In next sections, we develop mathematical descriptions of above stages and analyze these descriptions.

\section{Theory and Calculations}

In quantum mechanics, coherent states are special kinds of quantum states that their dynamics are very close to their corresponding classical system. Type of coherent state is used in problem, depends on the operators’ symmetry in Hamiltonian. Due to the  operators symmetry in Hamiltonian (1), For  more detailed  description and considering  all   multipolar  excitation  of coherent  states we used from coherent states in SU(3) group.  Vacuum state of this group is $(1,0,0)^T$  and coherent state for a single site in this group is introduced as [10]:
\begin{eqnarray}
|\psi \rangle &=& D^{1}(\theta, \phi)e^{-i\gamma \hat S^z} e^{2ig\hat Q^{xy}}|0\rangle \nonumber\\
\end{eqnarray}

where $D^{1}(\theta, \phi)$ is wigner function and $Q^{xy}$  is quadrupole moment. 
Two angles, $\theta$  and $\phi$ , the Euler angles, determine the direction of classical spin vector in spherical coordinate system. The angle $\gamma$  determines the direction of quadrupole moment around the spin vector and parameter g shows change of magnitude of spin vector and Average value of quadrupole  moment . Two angles $\phi$ and $\gamma$ change between 0 to $\pi$  and angle $\theta$  changes between $-\pi$ to $\pi$.

In order to derive Lagrangian from path integral, we drive the path integral from the following transition amplitude:

\begin{eqnarray}
P(\psi_1,t_1;\psi,t)=\langle \psi_1 | exp(-\frac{i}{\bar h}(\hat H(t_1-t)))| \psi \rangle
\end{eqnarray}
Using completeness relation and doing some mathematical work, the following equation for transition amplitude obtained:

\begin{eqnarray}
P(\psi_1,t_1;\psi,t)&=&lim_{n \rightarrow \infty}\sum_j \int D\mu_j(\psi) \nonumber\\
& &\times exp(\frac{-i}{\bar h}\int_t^{t_1}L_j(\theta,\phi,g,\gamma)d\tau) \nonumber\\
\end{eqnarray}

In the above equation L is Lagrangian and have the following form:[10]

\begin{eqnarray}
L=\bar h cos2g(cos\theta \phi_t +\gamma_t )-H(\theta, \phi, g, \gamma)
\end{eqnarray}

 $x_t=\frac{\partial x}{\partial t}$( $x=\theta, \phi$ ) and $H(\theta,\phi,g,\gamma)$ is classical Hamiltonian. By use of equations (4) and (5) and the action statationary principle, the classical equations of motion obtained. 

When  Lagrangian is obtained from path integral, another two terms  appear, one is kinetic term and has "Berry phase" properties that important in spin tunneling phenomena and the other depends on boundary condition values. In this paper, we do not consider these two terms.

Now classical equivalent of spin vector and their products must be computed so that the classical equivalent of Hamiltonian (1) is obtained. So consider:

\begin{eqnarray}
\vec S=\langle\psi|\vec {\hat S}|\psi\rangle
\end{eqnarray}

 as classical spin vector, and also consider: 

\begin{eqnarray}
\hat Q^{ij}=\frac{1}{2}(\hat S_i \hat S_j+\hat S_j \hat S_i -\frac{4}{3}\delta_{ij}I)
\end{eqnarray}

components of quadrupole moment. Because we can write any coherent state as product of single site coherent states, namely:

\begin{eqnarray}
|\psi \rangle=\prod_i |\psi \rangle_i
\end{eqnarray}

Then Spin operators in ground state of non-single ions Hamiltonian  can be commuted in different lattices [10]; so  

\begin{eqnarray}
\langle\psi|\hat S_{n}^i \hat S_{n+1}^j|\psi\rangle= \langle\psi | \hat S_n^i |\psi \rangle  \langle\psi | \hat S_{n+1}^j |\psi \rangle 
\end{eqnarray}

where $ |\psi \rangle=|\psi\rangle_n | \psi\rangle_{n+1}$.

The average spin values in SU(3) group for Coherent states (2) are defined as [16]:
\begin{eqnarray}
S^+ &=& e^{i\phi}cos(2g)sin\theta  \nonumber\\
S^- &=& e^{-i\phi} cos(2g)sin\theta \nonumber\\
S^z &=& cos(2g) cos\theta \nonumber\\
\end{eqnarray}

And also

\begin{eqnarray}
S^2&=& cos^2 (2g) \nonumber\\
q^2&=&sin^2(2g) \nonumber\\
S^2&+&q^2=1
\end{eqnarray}

In above relation, $S^2$ is related to dipole moment and $ q^2$ is related to quadrupole moment. Classical Hamiltonian can be obtained from the average calculation of Hamiltonian (1) over coherent states. The classical continuous limit of Hamiltonian in SU(3) group is:

\begin{eqnarray}
H&=&-\int \frac{dx}{a_0}(J cos^2( 2g)+K cos^4( 2g) \nonumber\\
&-  &\frac{a_0^2}{2}(4g_x^2 sin^2 (2g)(J+2Kcos^2 (2g))+cos^2 (2g)(J+2K cos^2 (2g))(\theta_x^2+\phi_x^2 sin^2\theta))) \nonumber\\
& &
\end{eqnarray}
Where $a_0$ is length of crystal sites.
The above classical Hamiltonian is substituted in equation of motion that obtained from the Lagrangian, and the result is classical equations of motion:

\begin{eqnarray}
\frac{1}{\omega_0}\theta_t&=&- a_0^2 cos(2g)(J+K+Kcos(4g)) \nonumber\\
& &\times \phi_{xx} sin\theta \nonumber\\
\frac{1}{\omega_0}\phi_t&=& a_0^2cos(2g)(J+K+Kcos(4g))(\phi_x^2 cos\theta \nonumber\\
& &+\theta_{xx} csc\theta) \nonumber\\
\frac{1}{\omega_0}g_t&=&0\nonumber\\
\frac{1}{\omega_0}\gamma_t&=&4 cos(2g)(J+K+Kcos(4g)) \nonumber\\
& &+(Kcos^3 (2g)(16g_x^2-8\theta_x^2 -5\phi_x^2 \nonumber\\
& &+3\phi_x^2 cos(2\theta)-2\theta_{xx})+cos(2g) \nonumber\\
& &\times (8g_x^2(J-K)-2J\theta_x^2 +8g_x^2Kcos(4g) \nonumber\\
& &+\frac{1}{2}J\phi_x^2 (-3+cos(2\theta))-J\theta_{xx}cot\theta) \nonumber\\
& &+4g_{xx}Jsin(2g)+2g_{xx} K(sin(2g) \nonumber\\
& &+sin(6g)))a_0^2 \nonumber\\
\end{eqnarray}

In these equations, $\omega_0$ is $\bar ha_0$. These equations describe nonlinear dynamics of non-Heisenberg ferromagnetic chain completely. If we omitted quadrupole excitation ($g=0$) in above equations, these equations reduced to Landau-Lifshitz equations. Then in comparission with Landau-Lifshitz equations, these equations are more complete and contain more degrees of freedom. Note that, solutions of these equations are different forms of magnetic solitons.

In this paper, only the linearized form of equations (13) for small excitation above the ground states is considered. To this end,at first, classical ground states must be calculated so in above Hamiltonian only non-derivative part is considered:

\begin{eqnarray}
H_0=-\int \frac{dx}{a_0}(Jcos^2 2g+Kcos^4 2g)
\end{eqnarray}

To find the smallest value of the $H_0$  we vary it respect to all the parameter, the ground state is obtained at:

\begin{eqnarray}
g=0, & & {   }g=\frac{\pi}{2}
\end{eqnarray}

In this paper, only dispersion of spin wave in neighborhood of the ground states is studied.
For this purpose, small linear excitations from the ground states, as shown in eq. (15), are defined:

\begin{eqnarray}
2g\rightarrow \pi+g
\end{eqnarray}

In this situation, the linearized classical equations of motion are:

\begin{eqnarray}
\frac{1}{\omega_0}\theta_t&=&- a_0^2(J+2K)\phi_{xx}  \nonumber\\
\frac{1}{\omega_0}\phi_t&=& a_0^2(J+2K)\theta_{xx}  \nonumber\\
\frac{1}{\omega_0}g_t&=&0\nonumber\\
\frac{1}{\omega_0}\gamma_t&=&-4(J+2K) \nonumber\\
\end{eqnarray}

Consider functions $\theta$ and $\phi$  as plane waves to obtain dispersion equation:

\begin{eqnarray}
\phi &=& \phi_0 e^{i(\omega t-kx)}+\bar {\phi_0}e^{-i(\omega t-kx)} \nonumber\\
\theta &=& \theta_0 e^{i(\omega t-kx)}+\bar {\theta_0}e^{-i(\omega t-kx)} \nonumber\\
\end{eqnarray}
 
Substitution of these equations in eq. (17), result in dispersion equation of spin wave near the ground states.

\begin{eqnarray}
\omega_1^2 &=& (J+2K)^2a_0^2 k^4 \omega_0^2 \nonumber\\
\omega_2 &=&-4 (J+2K)\omega_0 
\end{eqnarray}

It is evident from equation (19) that in addition to the dispersion acoustic branch, there exist non-dispersion optical branches which is related to the dipole and quadrapole excitations.

\section{Conclusion}

In this paper, describing equations of one-dimensional isotropic non-Heisenberg Hamiltonians are obtained using real-parameter coherent states. It is shown that both dipole and quadrupole excitations have different dispersion if there is small linear excitation from the ground state.

In addition, it is shown that for isotropic ferromagnets, the magnitude of average quadrupole moment is constant ($g_t= 0$) and its dynamics, is rotational dynamics around the classical spin vector ($\gamma_t \ne 0$).


\begin{thebibliography}{9}
\bibitem{law}	N. Manton and P. Sutcliffe, topological solitons,  Cambridge university press , (2004).
\bibitem{law}	E. L. Nagaev, Sov. Phys.,  25, 31 (1982); ´ E. L. Nagaev, Magnets with Nonsimple Exchange Interactions [in Russian], Nauka, Moscow (1988). 
\bibitem{law}V. M. Loktev and V. S. Ostrovski˘ı, Low Temp. Phys., 20,  (1994) , 775.
\bibitem{law} B. A. Ivanov, A. Yu. Galkin, R. S. Khymyn and A. Yu. Merkulov, Phys. Rev. B 77,  (2008), 064402.
\bibitem{law} V. S. Ostrovskii, Sov. Phys. JETP,  64(5),  (1986), 999.	
\bibitem{law}	A. Imambekov, M. Lukin, and M. Troyer, J. Phys. Rev. A 68, (2003), 063602.
\bibitem{law}	K. Harada, N. Kawashima, and M. Troyer, J. Phys. Soc. Japan 76, (2003), 013703.
\bibitem{law}	T. Grover and T. Senthil, Phys. Rev. Lett. 98, (2007) , 247202.
\bibitem{law}   N. Papanicolaou, Nuclear Physics B, 305 , (1988), 365.
\bibitem{law} Kh. O. Abdulloev, Kh. Kh. Muminov, Phys. Solid state,  36,  (1994).
 
\bibitem{law} Yu. A. Fridman, O. A. Kosmachev, and B. A. Ivanov, Phys. Rev. Lett. 106, (2011), 097202.
\bibitem{law} I. E. Dzyaloshinskii, Sol, St. Comm. 82, (1992), 579.
\bibitem{law} A. Garg,  Phys. Rev. B 67, (2003), 054406.
\bibitem{law} M. S. Foss-Feig and Jonathan R. Friedman, EPL, 86 (2009) 27002 .
\bibitem{law} M.Matusiewicz, M.Czerwinski, J.Kasperczyk and I.V.Kityk. Description of spin interactions in model $Fe_6S_6$ supercluster. Journ. Chem.Physics. V. 111, N 14, (1999), pp.6446-6455;
\bibitem{law} V G Makhankov et al, J. Phys. A: Math. Gen. 29 (2005).
\end{thebibliography}
\end{document}